\documentclass[aps,twocolumn,superscriptaddress,pra]{revtex4-1}
\usepackage{graphicx}  % needed for figures
\usepackage{dcolumn}   % needed for some tables
\usepackage{amssymb}   % for math

\usepackage{soul}   %yellow comments
\usepackage{graphicx}		% Include figure files
\usepackage{dcolumn}		% Align table columns on decimal point
\usepackage{bm}				% bold math

\usepackage[usenames,dvipsnames]{xcolor}
\usepackage{microtype}
\usepackage{times}
\usepackage{changes}
\usepackage{amsmath}
\usepackage{hyperref}
\hypersetup{
  colorlinks,
  allcolors=NavyBlue,
  linktoc=all
}

\newcommand*\dagg{^{\dagger}}

\newcommand*\nf{\bar{n}_f}

\usepackage{ulem}

\newcommand{\unit}[1]{\ensuremath{\,\mathrm{#1}}}
\newcommand{\s}[1]{\ensuremath{{}_\mathrm{#1}}}	% roman math subscript

\begin{document}

\preprint{APS/123-QED}

\title{Laser cooling of a nanomechanical oscillator  to the zero-point energy}

\author{Liu Qiu}\thanks{These two authors contributed equally.}
\author{Itay Shomroni}\thanks{These two authors contributed equally.}
\affiliation{Institute of Physics, \'Ecole Polytechnique F\'ed\'erale de Lausanne, Station 3, CH-1015 Lausanne, Switzerland}
\author{Paul Seidler}
\email{pfs@zurich.ibm.com}
\affiliation{IBM Research -- Zurich, S\"{a}umerstrasse 4, CH-8803 R\"{u}schlikon, Switzerland}
\author{Tobias J. Kippenberg}
\email{tobias.kippenberg@epfl.ch}
\affiliation{Institute of Physics, \'Ecole Polytechnique F\'ed\'erale de Lausanne, Station 3, CH-1015 Lausanne, Switzerland}

\date{\today}

\begin{abstract}
Optomechanical cavities in the well-resolved-sideband regime are ideally suited for the study of a myriad of quantum phenomena with mechanical systems, including backaction-evading measurements, mechanical squeezing, and generation of non-classical states.
For these experiments, the mechanical oscillator should be prepared in its ground state; residual motion beyond the zero-point motion must be negligible. The requisite cooling of the mechanical motion can be achieved using the radiation pressure of light in the cavity by selectively driving the anti-Stokes optomechanical transition.
To date, however, laser-absorption heating of optical systems far into the resolved-sideband regime
 has prohibited strong driving.  For deep ground-state cooling, previous studies have therefore resorted to passive cooling in dilution refrigerators.
Here, we employ a highly sideband-resolved silicon optomechanical crystal in a $^3$He buffer gas environment at $\sim\! 2\unit{K}$  to demonstrate laser sideband cooling to a mean thermal occupancy of $0.09_{-0.01}^{+0.02}$ quantum (self-calibrated using motional sideband asymmetry), which is $-7.4\unit{dB}$ of the oscillator's zero-point energy and corresponds to 92\% ground state probability.
Achieving such low occupancy by laser cooling opens the door to a wide range of quantum-optomechanical experiments in the optical domain.
\end{abstract}

\maketitle
Laser cooling techniques developed several decades ago~\cite{Chu1998,Cohen-Tannoudji1998,Wieman1999,Kippenberg2008} have revolutionized many areas of science and technology, with systems
ranging from atoms, ions and molecules~\cite{Ashkin1978,Wineland1979,Shuman2010,Anderegg2018,Ospelkaus2011,Monroe2013,Blatt2012} to solid-state structures and macroscopic objects~\cite{Schliesser2006,Li2011,LIGO2009}.
Among these systems, mechanical oscillators play a unique role given their macroscopic nature and their ability to couple to diverse physical quantities~\cite{Aspelmeyer2014}.
Laser cooling of  mechanical systems occurs via coupling of mechanical and electromagnetic degrees of freedom (optomechanical coupling) and has been demonstrated with a wide range of structures ~\cite{Wilson-Rae2007,Marquardt2007,Schliesser2006,Schliesser2008,Chan2011,Teufel2011,Verhagen2012,Wilson2015,Peterson2016,Clark2017,Rossi2018}.
It has led to the observation of radiation pressure shot noise~\cite{Purdy2013},
ponderomotive squeezing of light~\cite{Safavi-Naeini2013,Purdy2014},
and motional sideband asymmetry~\cite{Wilson-Rae2007,Safavi-Naeini2012,Weinstein2014,Sudhir2017,qiu_floquet_2019}.

Many optomechanical protocols, including mechanical squeezing~\cite{Kronwald2013,Lecocq2015,Wollman2015,Pirkkalainen2015}, entanglement~\cite{Ockeloen-Korppi2018}, state swaps~\cite{palomaki2013}, generation of non-classical states~\cite{Riedinger2016,Hong2017,Aspelmeyer2018,shomroni_optomechanical_2019}, and back-action evading (BAE) measurements below the standard quantum limit (SQL)~\cite{Clerk2008,suh2014,shomroni2019}, require ground state preparation of a well-sideband-resolved system, where Stokes and anti-Stokes motional transitions can be driven selectively.
In this case, driving of anti-Stokes transitions can be efficiently applied to damp the motion and sideband cool the system.
The cooling limit is set by laser noise (classical or quantum) or by technical limitations, such as absorption heating, and determines the residual thermal noise.
For the case of squeezing or BAE measurements, the amount of cooling beyond half quantum (equivalent to the zero point energy) determines the amount of squeezing or the amount to which the SQL on resonance is surpassed.
Such deep ground-state preparation has been demonstrated in microwave optomechanical systems~\cite{Teufel2011}. In the optical domain, however, cooling below half quantum has so far only been achieved in systems with low sideband resolution, i.e. in the bad-cavity limit~\cite{Peterson2016} or using feedback cooling~\cite{Rossi2018}.

Silicon optomechanical crystals (OMCs)~\cite{Eichenfield2009,Chan2012} that couple an optical mode at telecommunication wavelengths and a co-localized mechanical mode at GHz frequencies exhibit several exceptional features, including
large vacuum coupling rates $\mathcal{O}(1\unit{MHz})$~\cite{Chan2012} as well as ultralong phonon lifetime~\cite{Painter2019}.
They have been employed in a wide range of experiments,
such as continuous quantum measurements~\cite{Chan2011,qiu_floquet_2019,shomroni2019}, and probabilistic preparation of quantum states~\cite{Riedinger2016,Hong2017,Aspelmeyer2018}.
The compatibility of these systems with planar nanofabrication technology and their scalability have motivated studies of optomechanical topological phenomena~\cite{Schmidt2015,Brendel2017}, frequency conversion~\cite{Fang2016} and coupling to superconducting qubits~\cite{Keller2017,Arrangoiz-Arriola2018}.
Yet despite these promising features, ground-state preparation of silicon OMCs has only been possible via passive cooling to milli-Kelvin temperatures in dilution refrigerators~\cite{Meenehan2014,Meenehan2015}.
Significant heating due to optical absorption---a consequence of the extremely small optical mode volume and inefficient thermalization~\cite{holland_analysis_1963}---has limited experiments to use of weak laser pulses~\cite{Riedinger2016,Hong2017,Aspelmeyer2018,Painter2019} and precluded continuous measurements~\cite{Meenehan2014,qiu_floquet_2019,shomroni2019}.

%%% EXPERIMENTAL

\begin{figure}
	\includegraphics[scale=1]{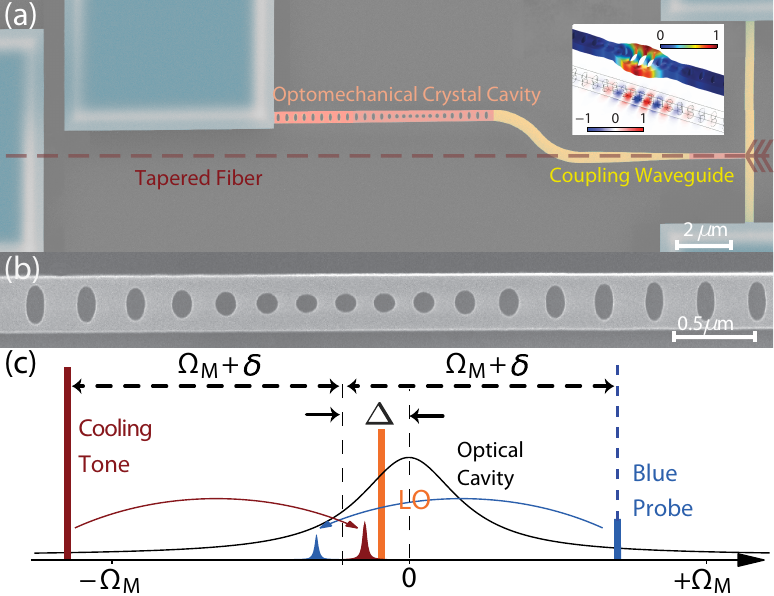}
	\caption{\textbf{Optomechanical crystal and experimental scheme.}
		(a)~False-color SEM image of the silicon optomechanical crystal cavity with a waveguide for input coupling of light.
		The path of the tapered fiber is indicated by the red dashed line.
		The inset shows the simulated mechanical breathing mode and optical mode.
		(b)~SEM image of the central portion of the silicon optomechanical crystal cavity.
		(c)~Measurement scheme using a cooling tone for sideband cooling and a blue probe for motional sideband asymmetry measurement.
		The local oscillator (LO) is used for detection and is not sent to the cavity.
	}
	\label{fig:exp}
\end{figure}

In this work, we demonstrate laser cooling of a strongly sideband-resolved silicon OMC to the zero-point energy, with residual mean phonon occupancy of $0.09_{-0.01}^{+0.02}$ (i.e. $-7.4\unit{dB}$ of the zero-point energy).
The measurement is self-calibrated using motional sideband asymmetry~\cite{Wilson-Rae2007,Weinstein2014,qiu_floquet_2019,Purdy2015,Underwood2015}.
Our experimental system, shown in Fig.~\ref{fig:exp}(a,b), consists of a quasi-one-dimensional silicon optomechanical crystal~\cite{Chan2012,qiu_floquet_2019,shomroni2019}.
The OMC is mounted in a $^3$He cryostat (Oxford Instruments HelioxTL) operated at $\sim 2.0\unit{K}$
and a buffer-gas pressure of $\sim 40\unit{mbar}$, which ensures efficient thermalization of the device~\cite{qiu_floquet_2019,shomroni2019}.
A tapered optical fiber is used to couple light evanescently into the coupling waveguide (40\% efficiency in this work).
For characterization, we monitor the laser light reflected from the single-sided optical cavity.
The optical resonance is at $1540\unit{nm}$ with a total linewidth of $\kappa/2\pi \simeq255\unit{MHz}$,
of which the external coupling rate is $\kappa\s{ex}/2\pi \simeq 71\unit{MHz}$.
The optical mode is coupled to a localized mechanical mode with frequency $\Omega_m/2\pi\simeq5.17\unit{GHz}$
and an intrinsic damping rate of $\Gamma\s{int}/2\pi \simeq 65\unit{kHz}$. The measured vacuum optomechanical coupling rate is $g_0/2\pi\simeq1080\unit{kHz}$.
The buffer-gas causes additional damping, increasing the mechanical linewidth to
$\Gamma_m=\Gamma\s{int}+\Gamma\s{gas}\simeq 2\pi\times 115\unit{kHz}$~\cite{suppmat}.

Motional sideband asymmetry, a signature of the quantum nature of the optomechanical interaction, was recently observed in various optomechanical systems~\cite{Safavi-Naeini2012,Weinstein2014,Sudhir2017}
and used to perform self-calibrated thermometry of the mechanical oscillator close to its ground state~\cite{Purdy2015,Underwood2015,qiu_floquet_2019}.
In our experiments, we adopt a two-tone pumping scheme [Fig.~\ref{fig:exp}(c)], where
a strong cooling tone near the lower motional sideband is applied for sideband cooling,
while an additional weaker `blue probe' is applied near the upper motional sideband.
By measuring the resonantly-enhanced anti-Stokes and Stokes scattered sidebands, proportional to $\bar{n}_f$ and $\bar{n}_f+1$, respectively, the mean phonon occupancy of the oscillator $\bar{n}_f$ can be determined.
The frequencies of the two tones are separated by $2(\Omega_m+\delta)$,
%where in our measurements $\delta<0$,
and their mean is detuned from the optical resonance frequency by $\Delta$ [Fig.~\ref{fig:exp}(c)].
In the presence of the cooling tone and blue probe, the mechanical susceptibility is modified by the radiation pressure.
The effective mechanical damping rate becomes $\Gamma\s{eff}=\Gamma_m+\Gamma\s{opt}$, with the total optomechanical damping rate (in the resolved-sideband regime) $\Gamma\s{opt}=-\Gamma_b+\Gamma_c$, where
\begin{equation}
\Gamma_{b(c)} = \bar{n}_{b(c)} g_0^2 \left(\frac{\kappa}{\kappa^2/4+(\Delta\pm\delta)^2}\right)
\label{eq:GammaEff}
\end{equation}
and $\bar{n}_b$ and $\bar{n}_c$ are the intracavity photon numbers of the blue probe and cooling tone, respectively.
In the weak coupling regime, $\Gamma\s{opt}\ll\kappa$, the effective mechanical frequency is $\Omega\s{eff}=\Omega_m+\delta\Omega_m$, with
\begin{equation}
\delta\Omega_m =  g_0^2 \left(\bar{n}_{b} \frac{\Delta+\delta}{\kappa^2/4+(\Delta+\delta)^2} +\bar{n}_c \frac{\Delta-\delta}{\kappa^2/4+(\Delta-\delta)^2}\right).
\label{eq:OmegaEff}
\end{equation}
The mean final phonon occupancy is given by
\begin{equation}
\nf = \frac{\Gamma_m \bar{n}\s{th}+\Gamma_b}{\Gamma\s{eff}},
\label{eq:nf}
\end{equation}
where $\bar{n}\s{th}$ is the mean phonon occupancy due to the thermal environment.
Importantly, the second term in the numerator of Eq.~\eqref{eq:nf} corresponds to quantum backaction (QBA) heating due to resonant Stokes transitions from the blue probe [Fig.~\ref{fig:exp}(c)]. This is in contrast to off-resonant Stokes transitions from the cooling tone, which are completely negligible in the well-resolved sideband regime (here $\Omega_m/\kappa\simeq 20$) and set the quantum limit for sideband cooling~\cite{Wilson-Rae2007,Marquardt2007,Peterson2016,Clark2017}.
In our two-tone experiments, QBA heating due to the blue probe, $\Gamma_b/\Gamma\s{eff}$, is comparable to the heating by the thermal bath at high probe powers and limits the cooling~\cite{suppmat}.
Thus we perform both two-tone measurements for ancillary quantum thermometry and single-tone measurements to achieve maximum cooling power.

In each measurement, we first determine the detuning of the cooling tone from the cavity, $\Delta_c=\Delta-\Omega_m-\delta$ by performing a coherent cavity response measurement~\cite{Safavi-Naeini2011,shomroni2019}.
We then obtain the thermomechanical noise spectrum by measuring the cavity output field using quantum-limited balanced heterodyne detection (BHD)
with a strong phase-locked local oscillator (LO; $\sim 8\unit{mW}$).
The frequency difference between the LO and the mean frequency of the two pumping tones is $\Delta\s{LO}$, where $0<-\delta<\Delta\s{LO}$.
The measured heterodyne noise spectrum, normalized to the shot noise floor, is given by
\begin{multline}
	   S_I(\Omega) = 1+\eta\Gamma\s{eff}\biggl[
	\frac{(\nf+1)\Gamma_b}{\Gamma\s{eff}^2/4+(\Omega+\delta-\Delta\s{LO})^2}	\\
	+\frac{\nf\Gamma_c}{\Gamma\s{eff}^2/4+(\Omega-\delta-\Delta\s{LO})^2}
	\biggr],
  \label{eq:SII}
\end{multline}
where $\eta$ is the overall detection efficiency.
The second and third terms in Eq.~\eqref{eq:SII} correspond to the scattered Stokes and anti-Stokes sidebands,
which we use for self-calibrated thermometry of the oscillator.
% Data1
\begin{figure*}[hbt!]
\includegraphics[scale=1]{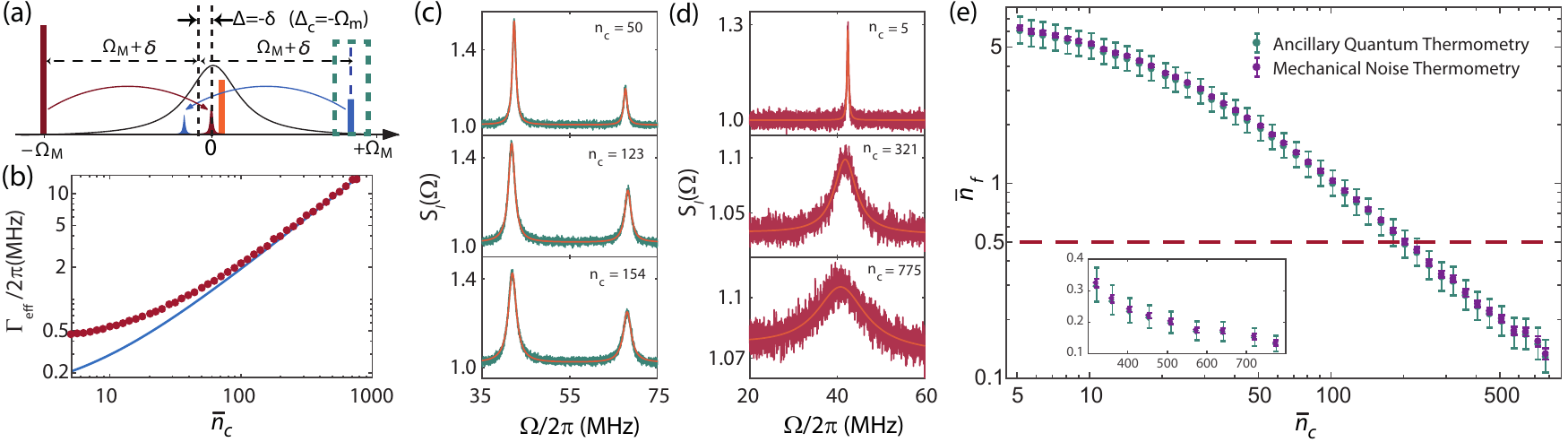}
\caption{
\textbf{Power dependence of sideband cooling.}
  (a)~Pumping scheme for the power sweep with a cooling tone at a fixed detuning of $-\Omega_m$ relative to the cavity resonance and an additional blue probe for sideband asymmetry calibration, as indicated in the dashed green box.
  The frequency separation between the cooling tone and blue probe is fixed at $2(\Omega_m+\delta)$.
  (b)~Measured effective mechanical linewidth $\Gamma\s{eff}$ from the noise power spectral density vs.~cooling tone intracavity photon number $\bar{n}_c$ (red full circles) in single-tone measurements with a theoretical plot with experimental values (blue curve).
  (c) and (d)~Single-sided noise spectra from balanced heterodyne detection normalized to the shot noise floor from two-tone and single-tone  measurements, respectively, with corresponding fit curves, for various intracavity photon numbers.
  (e)~Final phonon occupancy vs.~intracavity photon number of the cooling tone in single tone measurements.
  Purple open circles are anchored to the cryostat thermometer temperature at the lowest values of $\bar{n}_c$.
  Green full circles utilize the averaged calibration coefficient obtained from the ancillary two-tone sideband-asymmetry measurements,
  where the error bars are given by both the errors in the Lorentzian fit and in the calibration coefficient.
  The inset shows an expanded view at the highest cooling powers.
 The horizontal red dashed line corresponds to $\bar{n}_f=1/2$.
    }
\label{fig:nc}
\end{figure*}

Our scheme differs from previous experiments that utilize equal red and blue probes alongside a cooling tone~\cite{Weinstein2014,qiu_floquet_2019}.
By using only two tones, we avoid coupling between scattered sidebands due to Floquet dynamics that may introduce errors in the inferred phonon occupancy~\cite{qiu_floquet_2019}.
We keep the ratio between the input powers of the cooling tone and the blue probe around 6, to achieve both sufficient cooling and a measurable anti-Stokes signal ($\propto\bar{n}_f$).
From a series of two-tone measurements, we obtain a mean calibration coefficient between the normalized thermomechanical sideband area $A_c/\Gamma_c$ and the phonon occupancy $\nf$ using Eq.~\eqref{eq:SII}, where $A_c$ is the area of the sideband from the cooling tone~\cite{suppmat}.
The calibration coefficient serves as an ancillary quantum thermometer for the mechanical mode, independent of the resistive thermometer mounted in the cryostat.
For ground state cooling, we turn off the blue probe and perform single-tone sideband cooling measurements, keeping  the same experimental conditions and calibration. From the measured thermomechanical noise spectrum, we can thus obtain the final occupancy using two independent calibrations, \textit{i.e.}, the ancillary quantum thermometry and the mechanical noise thermometry, where for the latter the mechanical mode temperature is referenced to the cryostat thermometer~\cite{suppmat}.

% Data2
\begin{figure*}
  \includegraphics[scale=1]{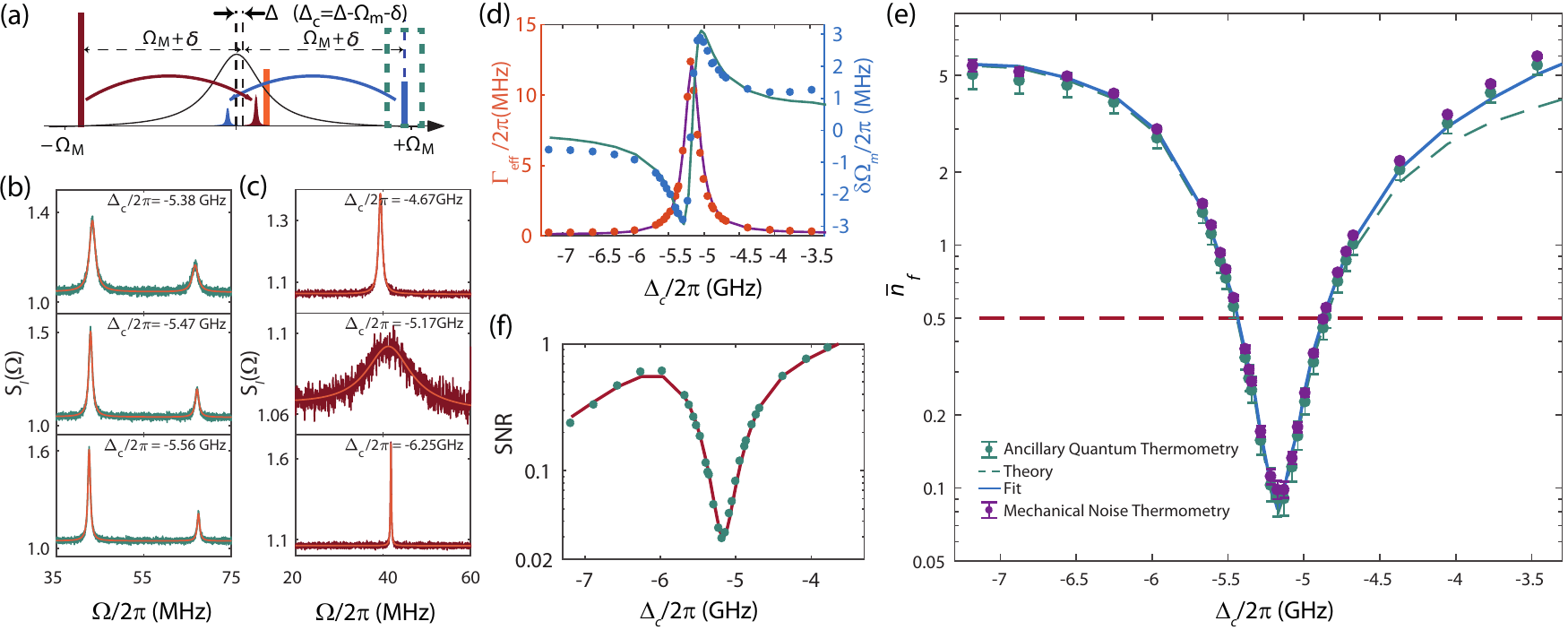}
  \caption{
  \textbf{Detuning dependence of the sideband cooling.}
  (a)~Pumping scheme for the detuning sweep where the detuning $\Delta_c$ of the cooling tone relative to the cavity resonance is varied.
  An additional blue probe is used for ancillary sideband asymmetry calibration, as indicated in the dashed green box.
  Frequency separation between the cooling tone and blue probe is fixed at $2(\Omega_m+\delta)$.
  (b) and (c)~Single-sided noise spectra from the balanced heterodyne detection normalized to the shot noise floor from two-tone and single-tone measurements, respectively, with corresponding fit curves, for various detunings.
  (d)~The fitted mechanical linewidth (red full circles, left axis) and optical spring effect (blue full circles, right axis), with corresponding theoretical plots based on experimental optomechanical parameters.
  (e)~Final phonon occupancy vs.~$\Delta_c$ in single-tone measurements.
  Green full circles show calibration using the ancillary two-tone quantum thermometry.
 Dashed green curve shows a theoretical plot calculated from experimental optomechanical parameters assuming ideal thermalization.
Blue curve shows a fitting curve incorporating excess optical heating.
  Error bars are given by the errors in the Lorentzian fits and in the calibration coefficient in two-tone sideband asymmetry.
  Purple full circles are anchored with cryostat thermometer at $\Delta_c/2\pi\simeq -7.18\unit{GHz}$.
    The horizontal red dashed line corresponds to $\bar{n}_f=1/2$.
  (f)~Signal-to-noise ratio (SNR) vs.~$\Delta_c$, with the fitting curve to a theoretical model which includes optical heating, with excess heating rate and overall detection efficiency as free fitting parameters.
}
\label{fig:delta}
\end{figure*}

In a first set of measurements shown in Fig.~\ref{fig:nc}, we vary the power of the pump tones while keeping $\Delta_c=-\Omega_m$ fixed for optimal sideband cooling.
A blue probe, as  indicated in the dashed green box in Fig.~\ref{fig:nc}(a), is utilized only for ancillary sideband asymmetry measurements.
Figure~\ref{fig:nc}(b) shows the effective mechanical linewidth $\Gamma\s{eff}$ as a function of the cooling-tone intracavity photon number $\bar{n}_c$, obtained from the noise spectra in the single-tone experiments (red full circles)
with a theoretical plot (blue curve) assuming a mechanical linewidth $\Gamma_m/2\pi=115\unit{kHz}$ and vacuum coupling rate $g_0/2\pi=1080\unit{kHz}$.
As shown in Fig.~\ref{fig:nc}(b), $\Gamma\s{eff}$ deviates from the theoretical value for low intracavity photon numbers. We attribute this to condensed $^3$He on the surface of OMC, which degrades the mechanical linewidth at low powers but may improve the thermalization~\cite{suppmat}.
Figure~\ref{fig:nc}(c)~and~(d) show a series of noise spectra from the two-tone and single-tone measurements, respectively, at various values of $\bar{n}_c$ along with Lorentzian fits.
The noise spectra are normalized to the shot noise floor, obtained by blocking the signal beam in the BHD.
The left and right thermomechanical sidebands shown in Fig.~\ref{fig:nc}(c) are due to the cooling tone and the blue probe, respectively.
We choose a series of pumping powers that ensures both sufficient laser cooling and measurable, non-overlapping Stokes and anti-Stokes sidebands~\cite{Weinstein2014}.
As the power increases, the ratio of the areas of the red and blue sidebands, given by $\nf\Gamma_c/(\nf+1)\Gamma_b$ [cf.~Eq.~\eqref{eq:SII}], decreases
as the mechanical oscillator approaches the ground state ($\nf\rightarrow 0$), as shown in Fig.~\ref{fig:nc}(c).
We thus obtain an averaged calibration coefficient between the normalized thermomechanical sideband area and the final occupancy from a series of ancillary quantum thermometry measurements~\cite{suppmat}.
At high pumping powers, we observe an increase in the noise floor, as evident in the middle and bottom panels in Fig.~\ref{fig:nc}(d). This originates from beating of the high signal power with excess noise of the LO around $5.17\unit{GHz}$~\cite{suppmat}.
Figure~\ref{fig:nc}(e) shows the inferred mean phonon occupancy $\nf$ vs.~$\bar{n}_c$  from the single-tone measurements, calibrated using two independent methods.
The green circles show the phonon occupancy calibrated using the ancillary sideband asymmetry measurements.
The purple open circles show the calibration using the cryostat thermometer by anchoring the lowest value of $\bar{n}_c\sim 5$ at $2.0\unit{K}$.
This calibration requires knowledge of $\Gamma_m$, which is estimated by subtracting the calculated value of $\Gamma\s{opt}$ at this power from the measured $\Gamma\s{eff}$, to yield $\Gamma_m/2\pi\simeq 360\unit{kHz}$.
We note that $\Gamma_m$ is unnecessary using ancillary quantum thermometry, making it an ideal independent quantum thermometer, as opposed to conventional mechanical noise thermometry.
More details on the two different calibrations of $\nf$ are given in the Supplemental Material~\cite{suppmat}.
The two methods show excellent agreement.
The minimum phonon occupancy achieved in this power-sweep experiment is $0.13_{-0.02}^{+0.02}$ (88\% ground-state occupancy) and is reached at a cooling-tone intracavity photon number of $\bar{n}_c\approx 776$.

In a second set of measurements, we vary the detuning $\Delta_c$ of the cooling tone with respect to the cavity resonance, keeping the frequency separation of the blue probe in the ancillary measurement and that of the LO fixed at $2(\Omega_m+\delta)$ and $\Delta\s{LO}$, respectively [Fig.~\ref{fig:delta}(a)].
Figure~\ref{fig:delta}(b) and (c) each show a  series of measured noise spectra normalized to the shot noise floor at various values of $\Delta_c$, together with Lorentzian fits, from the two-tone and single-tone measurements.
In the ancillary two-tone measurements, the input powers of the cooling tone and blue probe are $\sim 350\unit{\mu W}$ and $\sim  60\unit{\mu W}$ respectively, with a series of values of $\Delta_c$ to ensure sufficient laser cooling and measurable, non-overlapping Stokes and anti-Stokes sidebands.
To infer $\nf$ via sideband thermometry, the detuning-dependent intracavity photon number and optical susceptibility for the two scattered sidebands must be taken into consideration.
We obtain a mean calibration coefficient between the normalized thermomechanical sideband area and the final occupancy from the sideband asymmetry measurements.
For single-tone measurements, the cooling tone input power is $\sim 500\unit{\mu W}$.
Figure~\ref{fig:delta}(d) shows the effective mechanical linewidth (red circles) and the optical spring effect (blue circles) obtained from a Lorentzian fit to the noise spectrum, with excellent agreement with the respective theoretical curves.
We note that, due to the presence of high input power throughout the measurement, the mechanical linewidth degradation observed at low powers in the previous measurement [Fig.~\ref{fig:nc}(b)] is absent.
Figure~\ref{fig:delta}(e) shows $\nf$ vs.~$\Delta_c$, where $\nf$ is calibrated from the thermomechanical sideband area from the single-tone sideband cooling measurements.
Green circles are determined using the mean calibration factor obtained from sideband asymmetry measurements.
The theoretical dependence calculated from experimental parameters~\eqref{eq:nf} is shown as a green dashed curve for comparison.
The theory curve is in excellent agreement with the data except in the region where the cooling tone approaches the cavity resonance, indicating residual optical heating~\cite{Meenehan2014,qiu_floquet_2019}.
We fit the phonon occupancy with a model incorporating heating [blue curve in Fig.~\ref{fig:delta}(e)] that is both linear and quadratic in the number of intracavity photons~\cite{suppmat}.
The fit indicates that the excess optical heating in our measurements has primarily a quadratic dependence, resulting in an increase in $\bar{n}\s{th}$ of $\sim1.2\times10^{-6}\bar{n}_c^2$; the linear coefficient is negligible.
This is different from previous experiments with large optical decay rate, where linear absorption heating dominates~\cite{qiu_floquet_2019}.
The quadratic dependence is suggestive of two-photon-absorption~\cite{barclay_nonlinear_2005,pernice_carrier_2011}.
We note that in any case such optical heating cannot come from excess laser noise~\cite{Rabl2009,jayich_cryogenic_2012,safavi-naeini_laser_2013,kippenberg_phase_2013}, for which the heating rate peaks at $\Delta_c=-\Omega_m$.
For the noise thermometry, we anchor the calibration to $2.0\unit{K}$, at farthest detuning of $\Delta_c/2\pi\approx-7.18\unit{GHz}$. The resulting data are shown as purple full circles in Fig.~\ref{fig:delta}(e).
For $\Delta_c/2\pi=-7.18\unit{GHz}$ with $\bar{n}_c=330$, the estimated increase in $\bar{n}\s{th}$ due to quadratic heating is $\sim 0.135$, which is negligible compared to the bare thermal bath occupation of 8.2 phonons.
This indicates that the mechanical oscillator is well-thermalized despite the high pumping power.
The minimum phonon occupancy, occurring close to the red mechanical sideband ($\Delta_c/2\pi\simeq -5.17\unit{GHz}$), is $\nf = 0.09_{-0.01}^{+0.02}$, which is $-7.4\unit{dB}$ of the zero-point energy.
This places the mechanical oscillator at 92\% ground state occupation.
In Fig.~\ref{fig:delta}(f), the signal-to-noise ratio vs.~$\Delta_c$ for the thermomechanical noise spectrum is shown with a fit~\cite{suppmat} that includes the quadratic heating model in addition to the standard sideband cooling theory.  The fit yields an overall detection efficiency $\eta\simeq 6.4\%$.

In conclusion,
we have demonstrated high-fidelity sideband cooling to the zero-point energy of a localized GHz mechanical mode of a silicon OMC. The residual mean phonon occupancy is $0.09_{-0.01}^{+0.01}$ (92\% ground state occupation).
The system possess a unique blend of advantageous properties, combining high mechanical frequency, large sideband resolution, negligible optical-absorption heating and the ability to be prepared in the ground state in the presence of strong probing.
These characteristics enable a large number of quantum optomechanical experiments that have remained elusive in the optical domain, including two-tone backaction-evading measurements reaching sub-SQL sensitivity~\cite{suh2014,shomroni2019,shomroni_two-tone_2019}, squeezed mechanical states~\cite{Kronwald2013,Wollman2015,Lecocq2015,Pirkkalainen2015,lei2016}, low-added-noise optomechanical transducers~\cite{Fang2016,Peterson2017,Bernier2017}, as well as quantum-coherent operations such as state swapping~\cite{palomaki2013} and entanglement generation~\cite{Ockeloen-Korppi2018,palomaki2013b}. %In addition, our results provide valuable insight relevant to eventual microwave-optical frequency conversion with piezoelectric optomechanical platforms.

\begin{acknowledgements}
This work is supported by the European Union’s Horizon 2020 research and innovation programme under grant No.~732894 (FET Proactive HOT).
Samples were fabricated in the Binnig and Rohrer Nanotechnology Center at IBM Research--Zurich and at the Center of MicroNanoTechnology (CMi) at EPFL.
\end{acknowledgements}

\section*{Data Availability Statement}

All data and analysis files will be made available via \texttt{zenodo.org} upon publication.

\makeatletter
\close@column@grid
\clearpage
%\twocolumngrid

\onecolumngrid
\begin{center}
	\textbf{\large
		\@title\\[0.5cm]
		Supplemental Material\\[.5cm]
	}
	Liu Qiu,$^{1,*}$ Itay Shomroni,$^{1,*}$ Paul Seidler,$^{2,\dagger}$ and Tobias J. Kippenberg$^{1,\ddagger}$\\[.1cm]
	{\itshape
		$^1$Institute of Physics, \'Ecole Polytechnique F\'ed\'erale de Lausanne, Lausanne CH-1015, Switzerland\\
		$^2$IBM Research -- Zurich, S\"{a}umerstrasse 4, CH-8803 R\"{u}schlikon, Switzerland
	}\\
	$^\dagger$Electronic address: pfs@zurich.ibm.com\\
	$^\ddagger$Electronic address: tobias.kippenberg@epfl.ch\\
	(\@date)\\[1cm]
\end{center}

\twocolumngrid

\makeatother

\setcounter{equation}{0}
\setcounter{figure}{0}
\setcounter{table}{0}
\setcounter{page}{1}
\renewcommand{\theequation}{S\arabic{equation}}
\renewcommand{\thefigure}{S\arabic{figure}}

\section{Theory}
\label{sec:theory}

In the sideband asymmetry experiments, we pump the optomechanical system with two tones, one close to the red motional sideband and the other close to the blue motional sideband of the cavity.
The amplitude of the input field takes the form $a\s{in} = a_c e^{-i\omega_c t}+ a_b e^{-i\omega_b t}+\delta a\s{in}$, where $a_{c(b)}$ and $\omega_{c(b)}$ are the amplitude and frequency of the cooling tone (blue probe), and $\delta a\s{in}$ corresponds to the input noise.
The two tones are separated by $2(\Omega_m+\delta)$, and the mean of their frequencies is detuned from the cavity resonance by $\Delta$, as shown in Fig. 1 in the main text.
The mechanical mode is coupled to the optical field through radiation pressure and is additionally coupled to a thermal reservoir.
By linearizing the intracavity optical field $a\rightarrow\bar{a}+\delta a$ and the mechanical displacement $b\rightarrow\bar{b}+\delta{b}$,
we obtain the quantum Langevin equations for the fluctuation of the intracavity fields in the frame rotating at the mean of the frequencies of the blue probe and the cooling tone~\cite{Aspelmeyer2014,Weinstein2014,Wollman2015}. Within the rotating-wave approximation,
\begin{align}
\delta\dot a &= \Bigl(i\Delta-\frac{\kappa}{2}\Bigr)\delta a + i(g_c\delta b+ g_b\delta b\dagg)+ \sqrt{\kappa\s{ex}} \delta a\s{in} + \sqrt{\kappa_0}\delta a\s{vac}\nonumber\\
\delta\dot b &= \Bigl(i\delta-\frac{\Gamma_m}{2}\Bigr)\delta b + i(g_c\delta a +g_b\delta a\dagg) +\sqrt{\Gamma_m}\delta b\s{in},
\label{eq:qlet}
\end{align}
where $g_b=g_0\sqrt{\bar{n}_b}$ and $g_c=g_0\sqrt{\bar{n}_c}$, $\bar{n}_{b(c)}$ is the intracavity photon number due to the blue probe (cooling tone) and $g_0$ is the vacuum optomechanical coupling rate.
$\kappa$, $\kappa_0$, and $\kappa\s{ex}$ are the total, intrinsic and external optical loss rates of the optical mode,
$\Gamma_m$ is the mechanical damping rate, and
$\delta a\s{in}$, $\delta a\s{vac}$, and $\delta b\s{in}$ correspond to the optical input noise, optical vacuum noise and the mechanical noise.
The optical and mechanical noise operators satisfy the following noise correlations,
\begin{align}
\langle\delta a\s{in}(t) \delta a\s{in}\dagg (t')\rangle & =\alpha \delta (t-t') \nonumber\\
\langle\delta a\s{in}\dagg(t) \delta a\s{in} (t')\rangle & = 0\nonumber\\
\langle\delta a\s{vac}(t) \delta a\s{vac}\dagg (t')\rangle & = \alpha \delta (t-t') \nonumber\\
\langle\delta a\s{vac}\dagg(t) \delta a\s{vac} (t')\rangle & = 0 \nonumber\\
\langle\delta b\s{in}(t) \delta b\s{in}\dagg (t')\rangle & = \bar{n}\s{th} \delta (t-t') \nonumber \\
\langle\delta b\s{in}\dagg(t) \delta b\s{in} (t')\rangle & =(\bar{n}\s{th}+\beta) \delta (t-t'),
\end{align}
where $\bar{n}\s{th}=k_B T/\hbar\Omega_m$ is the mean phonon occupation of the mechanical oscillator in equilibrium with the thermal reservoir at temperature $T$.
Here $\alpha$ describes the fluctuation in the optical field while $\beta$ describes the zero-point fluctuation in the mechanical motion.
In general, $\alpha$ (for a quantum limited laser) and $\beta$ equal to one.

Here we define the optical and mechanical susceptibility as
\begin{align}
\chi_c(\omega) & : = \frac{1}{\kappa/2 - i (\omega+\Delta)} \nonumber\\
\chi_m(\omega) & : = \frac{1}{\Gamma_m/2 - i (\omega+\delta)} \nonumber.
\end{align}
Solving Eq.~\eqref{eq:qlet} in the Fourier domain, we thus obtain
\begin{widetext}
	\begin{equation}
	\begin{split}
	\delta a &= \chi_c(\omega)[\sqrt{\kappa\s{ex}}\delta a\s{in}(\omega)+\sqrt{\kappa_0}\delta a\s{vac}(\omega)+i g_c \delta b+i g_b \delta b\dagg]
	\\
	\delta a\dagg &= \chi_c^*(-\omega)[\sqrt{\kappa\s{ex}}\delta a\s{in}\dagg(\omega)+\sqrt{\kappa_0}\delta a\s{vac}\dagg(\omega)-i g_c\delta b\dagg-i g_b\delta b]
	\\
	\begin{pmatrix}
	\delta b(\omega) \\ \delta b\dagg(\omega)
	\end{pmatrix} &=
	\frac{i}{N(\omega)}
	M(\omega)
	\left[
		\sqrt{\kappa\s{ex}}
		\begin{pmatrix} \delta a\s{in}\\ \delta a\s{in}\dagg \end{pmatrix}
		+ \sqrt{\kappa_0}
		\begin{pmatrix} \delta a\s{vac}\\ \delta a\s{vac}\dagg \end{pmatrix}
	\right]
	%\\ &\qquad
	+\frac{\sqrt{\Gamma_m}}{N(\omega)}
	\begin{pmatrix}
	{\chi_m^*}^{-1}(-\omega)-i\Sigma^*(-\omega)
	&-i\Pi(\omega)\\
	i \Pi(\omega)
	& \chi_m^{-1}(\omega)+i\Sigma(\omega)
	\end{pmatrix}
	\begin{pmatrix}
	\delta b\s{in}\\
	\delta b\s{in}\dagg
	\end{pmatrix},
	\end{split}
	\end{equation}
where
\begin{equation}
\begin{split}
M(\omega) &= \begin{pmatrix}
\chi_c(\omega) g_c ({\chi_m^*}^{-1}(-\omega)+G^2\chi_c^*(-\omega))
& \chi_c^*(-\omega) g_b ({\chi_m^*}^{-1}(-\omega)+G^2\chi_c(\omega))\\
\chi_c(\omega) g_b (\chi_m^{-1}(\omega)+G^2\chi_c^*(-\omega))
&\chi_c^*(-\omega) g_c (\chi_m^{-1}(\omega)+G^2\chi_c(\omega))
\end{pmatrix}\\
N(\omega) &= \chi_m^{-1}(\omega){\chi_m^*}^{-1}(-\omega)+i{\chi_m^*}^{-1}(-\omega)\Sigma(\omega)
-i\chi_m^{-1}(\omega)\Sigma^*(-\omega)
+G^4\chi_c(\omega)\chi_c^*(-\omega)
\end{split}
\end{equation}
\end{widetext}
and
\begin{gather}
\Pi(\omega) = -i g_c g_b [\chi_c(\omega)-\chi_c^*(-\omega)] \\
\Sigma(\omega) = -i[g_c^2\chi_c(\omega)-g_b^2 \chi_c^*(-\omega)] \\
G^2 = g_c^2 - g_b^2.
\end{gather}
The mechanical susceptibility, which is modified by the radiation pressure from the two tones, is defined as
\begin{equation}
\label{eq:chimmod}
\begin{split}
\chi\s{meff}(\omega) &= \frac{{\chi_m^*}^{-1}(-\omega)-i \Sigma^*(-\omega)}{N(\omega)} \\
&\approx \frac{1}{(\Gamma_m+\Gamma\s{opt})/2-i(\omega+\delta-\delta\Omega_m)}.
\end{split}
\end{equation}
During our measurements, the ratio of cooling-tone to blue-probe pumping powers is fixed around 6.
In the weak-coupling regime ($\Gamma\s{opt}\ll\kappa$), the effective damping rate of the mechanical oscillator becomes $\Gamma\s{eff}=\Gamma_{m}+\Gamma\s{opt}$, where the optomechancial damping rate (in the resolved-sideband limit) is $\Gamma\s{opt}=-\Gamma_b+\Gamma_c$, and $\Gamma_b$ and $\Gamma_c$ take the form
\begin{equation}
\Gamma_{b(c)} =\bar{n}_{b(c)} g_0^2 \left( \frac{\kappa}{\kappa^2/4+(\Delta\pm\delta)^2}\right)
\label{eq:Gamma}.
\end{equation}
The optical spring effect is given by
\begin{equation}
\begin{split}
\delta\Omega_m &= \bar{n}_b g_0^2 \left( \frac{\Delta+\delta}{\kappa^2/4+(\Delta+\delta)^2}\right) \\
&\qquad +\bar{n}_c g_0^2 \left( \frac{\Delta-\delta}{\kappa^2/4+(\Delta-\delta)^2}\right).
\end{split}
\end{equation}
From the Wiener-Khinchin theorem, the two-sided mechanical displacement noise spectrum is calculated in the lab frame as
\begin{equation}
\begin{split}
\frac{S_{xx}(\omega)}{x\s{zpf}^2}
&= S_{bb}(\omega)+S_{b\dagg b\dagg}(\omega) \\
&=\frac{\Gamma_m \left(\bar{n}\s{th}+1\right)+\Gamma_c}{(\omega-\Omega\s{eff})^2+\Gamma\s{eff}^2/4}
%\\ &\qquad
+\frac{\Gamma_m \bar{n}\s{th}+\Gamma_b}{(\omega+\Omega\s{eff})^2+\Gamma\s{eff}^2/4}.
\end{split}
\end{equation}
The final mechanical occupation, in the sideband resolved limit, is given by
\begin{equation}
\nf = \frac{\Gamma_m \bar{n}\s{th}+\Gamma_b}{\Gamma\s{eff}}.
\label{eq:nf_SI}
\end{equation}
In the two-tone pumping scheme, the quantum backaction (QBA) from the blue probe can become dominant even when there is no heating due to optical absorption,
as is evident from the second term in the numerator of Eq.~\eqref{eq:nf_SI}.

When coupled to both the optical and thermal reservoirs, the zero point fluctuation of the dressed mechanical mode  becomes~\cite{weinstein_quantum_2016}
\begin{equation}
\tilde{\beta} = \frac{\alpha \left(\Gamma_c - \Gamma_b\right)+\Gamma_m\beta}{\Gamma\s{eff}}.
\end{equation}
For $\alpha =1$ (i.e. a quantum limited laser field) and $\beta=1$, we see that also $\tilde{\beta} = 1$.

We note that Eq.~\eqref{eq:nf_SI} is formulated using the rotating-wave approximation, where the QBA from the cooling tone is neglected~\cite{Wilson-Rae2007,Marquardt2007,Clark2017}, as the system is deep in the resolved-sideband regime.
In the following, we explain this conclusion using a Raman-scattering picture that addresses QBA from both the cooling tone and the blue probe~\cite{Marquardt2007}.
Without the mechanical damping, the mean phonon occupancy of the optomechanical crystal cavity $\bar{n}\s{min}$ is given by the detailed balance expression
\begin{equation}
\frac{\bar{n}\s{min}+1}{\bar{n}\s{min}}=\frac{\Gamma^{AS}_b+\Gamma^{AS}_c}{\Gamma^S_b+\Gamma^S_c},
\end{equation}
where $\Gamma^{AS}_{b(c)}$ and $\Gamma^{S}_{b(c)}$ correspond to the anti-Stokes and Stokes scattering rate, respectively, of the blue probe (cooling tone).  Now, $\Gamma^{AS}_c\equiv\Gamma_c$ and $\Gamma^{S}_b\equiv\Gamma_b$ [Eq.~\eqref{eq:Gamma}], whereas $\Gamma^{AS}_b$ and $\Gamma^{S}_c$ take the form
\begin{align}
\Gamma^{AS}_b &=\bar{n}_b g_0^2 \left( \frac{\kappa}{\kappa^2/4+(\Delta+\delta+2\Omega_m)^2}\right) \nonumber\\
\Gamma^S_c &=\bar{n}_c g_0^2 \left( \frac{\kappa}{\kappa^2/4+(\Delta-\delta-2\Omega_m)^2}\right)
\label{eq:GammaRaman}.
\end{align}
The imbalanced Stokes and anti-Stokes scattering from both the cooling tone and the blue probe leads to a net optomechanical damping of the mechanical oscillator $\Gamma\s{opt}=\Gamma^{AS}_b+\Gamma^{AS}_c-\Gamma^S_b-\Gamma^S_c\approx\Gamma_c-\Gamma_b$.
The minimum phonon occupancy $\bar{n}\s{min}$ is therefore given by
\begin{equation}
\bar{n}\s{min}=\frac{\Gamma^S_c+\Gamma_b}{\Gamma\s{opt}}.
\label{eq:nmin}
\end{equation}
The stochastic QBA force from both tones produces a residual phonon occupancy of the optomechanical crystal cavity.
In the deep-resolved-sideband regime ($\kappa\ll\Omega_m$), such that $\Gamma^S_c\ll\Gamma\s{opt}$, the QBA from the cooling tone is negligible. After including the mechanical damping $\Gamma_m$, $\nf$ takes the form in Eq.~\eqref{eq:nf_SI}, where the QBA from only the blue probe is considered.

Adopting the standard input-output formalism, we can obtain the output optical field $\delta a\s{out} = \delta a\s{in}-\sqrt{\kappa\s{ex}}\delta a$.
To achieve a quantum-limited measurement of the output field, we use balanced heterodyne detection,
beating the reflected optical signal with a strong local oscillator.
The frequency difference between the local oscillator and the mean frequency of the two pumping tones is $\Delta\s{LO}$, where $0<-\delta<\Delta\s{LO}$.
The measured single-sided heterodyne noise spectrum corresponds to the symmetrized autocorrelator of the photocurrent,
$S_I(\Omega)=\frac12\int_{-\infty}^\infty\langle\{\overline{\hat{I}_{\text{out}}(t+t'),\hat{I}_{\text{out}}(t')}\}\rangle e^{i\Omega t}dt$, and, when normalized to the shot noise, is given by
\begin{multline}
	   S_I(\Omega) = 1+\eta\Gamma\s{eff}\biggl[
	\frac{(\nf+1)\Gamma_b}{\Gamma\s{eff}^2/4+(\Omega+\delta-\Delta\s{LO})^2}	\\
	+\frac{\nf\Gamma_c}{\Gamma\s{eff}^2/4+(\Omega-\delta-\Delta\s{LO})^2}
	\biggr],
\label{eq:SII_SI}
\end{multline}
where $\eta$ is the overall detection efficiency. In~\eqref{eq:SII_SI}, the first term corresponds to the shot noise, whereas the second and third terms correspond to the thermomechanical sidebands of the blue probe and cooling tone.

We use Eq.~\eqref{eq:SII_SI} to determine the phonon occupancy $\nf$ from the asymmetry of the motional sidebands, considering the detuning dependent scattering rate, Eq.~\eqref{eq:Gamma}.
We then obtain a calibration coefficient between the normalized thermomechanical sideband area ($A_c/\Gamma_c$) and the phonon occupancy, where $A_c$ is the sideband area under the Lorentzian noise spectrum.

To achieve high fidelity ground state preparation, we employ only the cooling tone to avoid the QBA heating from the blue probe. The final occupancy of the mechanical oscillator is $\nf = \Gamma_m \bar{n}\s{th}/\Gamma\s{eff}$, where $\Gamma\s{eff}=\Gamma_m+\Gamma_c$. When the optical absorption heating is present, the final occupancy becomes
\begin{equation}
  \label{eq:nfabs}
  \nf = \frac{(\bar{n}\s{th}+\alpha_1\bar{n}_c+\alpha_2\bar{n}_c^2)\Gamma_m}{\Gamma\s{eff}}.
\end{equation}
Here we assume optical absorption heating both linear and quadratic in $\bar{n}_c$ with coefficients $\alpha_1$ and $\alpha_2$ respectively.
The signal to noise ratio (SNR) of the mechanical sideband in the noise spectrum can be calculated,
\begin{equation}
\label{eq:SNR}
\mathrm{SNR} = 4 ( \bar{n}\s{th}+\alpha_1 \bar{n}_c+ \alpha_2 \bar{n}_c^2 ) \eta
\frac{ \bar{n}_c C_0 \frac{(\kappa/2)^2}{(\kappa/2)^2+(\Delta_c+\Omega_m)^2}}
{(\bar{n}_c C_0 \frac{(\kappa/2)^2}{(\kappa/2)^2+(\Delta_c+\Omega_m)^2}+1)^2 },
\end{equation}
where $C_0$ is the vacuum optomechanical cooperativity. For optimal detuning, $\Delta_c=-\Omega_m$, and high pumping powers, $\bar{n}_c C_0\gg 1$, we have $\mathrm{SNR}\simeq 4\eta\nf$, which depends only on the overall detection efficiency and the final occupancy~\cite{qiu_floquet_2019}.

\section{Fabrication}
\label{sec:fabrication}

Our OMCs are fabricated on a silicon-on-insulator wafer (Soitec) with a top-silicon device-layer thickness of $220\unit{nm}$ and a buried-oxide layer thickness of $2\unit{\mu m}$.
We pattern our chips by electron beam lithography using
4\% hydrogen silsesquioxane (HSQ) as a negative resist.
Pattern transfer into the device layer is accomplished by inductively-coupled-plasma reactive ion etching (ICP-RIE) with a mixture of HBr and O$_2$.
To permit input/output coupling with a tapered fiber, an additional photolithography step is performed followed by reactive ion etching (RIE) with a mixture of $\unit{SF_6}$ and $\unit{C_4 F_8}$ to create a mesa structure.
After resist removal, the buried oxide layer is partially removed in 10\% hydrofluoric acid
to create free-standing devices.
Following a Piranha (a mixture of sulfuric acid and hydrogen peroxide)
cleaning step to remove organic residues,
the sample is finally dipped into 2\% hydrofluoric acid to terminate the silicon surface with hydrogen atoms.
The chip is then immediately mounted on the sample holder for characterization and loaded into the cryostat.

\section{Experimental System Details}
\label{sec:exp_details}

\begin{figure}
	\includegraphics[scale=1]{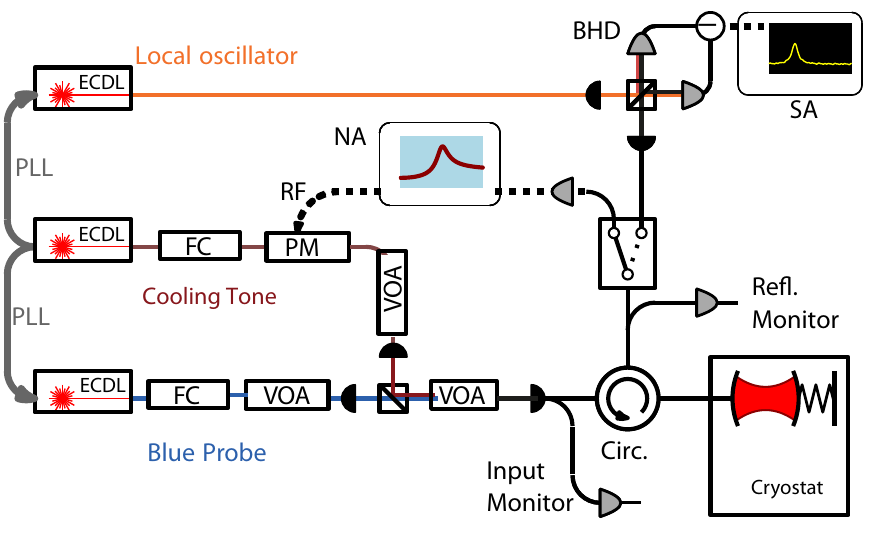}
	\caption{\textbf{Experimental setup.}
		ECDL, external-cavity diode laser; FC, filter cavity; PM, phase modulator; VOA, variable optical attenuator;
		BHD, balanced heterodyne detector; SA, spectrum analyzer; NA, network analyzer; PLL, phase-locked loop.}
	\label{fig:setup}
\end{figure}

The experiments are performed in a $^3$He buffer gas cryostat (Oxford Instruments HelioxTL)
capable of reaching a base temperature of $500\unit{mK}$.
The surrounding gaseous $^3$He improves the thermalization of the silicon OMC and thereby significantly diminishes the temperature increase due to optical absorption, as shown in previous optomechanical experiments~\cite{qiu_floquet_2019,shomroni2019}.
The optical resonance is at $1540\unit{nm}$ with a total linewidth of $\kappa/2\pi \simeq255\unit{MHz}$,
of which the external coupling rate is $\kappa\s{ex}/2\pi \simeq 71\unit{MHz}$.
The optical mode is coupled to a localized mechanical mode with frequency $\Omega_m/2\pi\simeq5.17\unit{GHz}$ with an intrinsic damping rate $\Gamma\s{int}/2\pi \simeq 65\unit{kHz}$.
An independent measurement is performed at temperature of $4\unit{K}$ and pressure of $40\unit{mbar}$, from which we obtain vacuum optomechanical coupling rate $g_0/2\pi=1.08\pm{0.006}\unit{MHz}$ and broadened mechanical damping rate $\Gamma_m/2\pi = 115\pm{8}\unit{kHz}$ due to additional gas damping .

In our experiment, we work at $2.0\unit{K}$ and $^3$He buffer gas pressure $\sim 40\unit{mbar}$.
A schematic of the experimental setup is shown in Fig.~\ref{fig:setup}.
Three external cavity diode lasers (ECDLs) generate the local oscillator (Toptica CTL 1550), cooling tone (Toptica CTL 1500), and blue probe (Toptica CTL 1500).
The blue probe and local oscillator (LO) are phased-locked to the cooling tone.
Both cooling tone and the blue probe are filtered by a $50\unit{MHz}$ bandwidth tunable Fabry-Perot filters which are locked to the respective tones using PDH lock technique, to reject the high frequency excess laser phase noise.
The cooling tone passes through a phase modulator (PM), used to generate weak sidebands as probes for coherent optomechanical spectroscopy.
The cooling tone and the blue probe are combined in free-space with the same polarization and sent into a single-mode fiber that enters the cryostat.
The single-pass coupling efficiency from the tapered fiber to the cavity input mirror is $\sim 40\%$.
A fiber-optic circulator feeds the reflected light to the detection stage, which can be toggled between two different paths.
In the first path, the reflected light is sent to a fast photoreceiver connected to a network analyzer for coherent optomechanical spectroscopy, in which case the phase modulator is employed.
In the second path, the reflected light is sent to a balanced heterodyne detection (BHD) setup, where it is mixed with a strong local oscillator ($\sim 8\unit{mW}$) on balanced photodetectors.
The power spectral density of the photocurrent is analyzed by a spectrum analyzer.
In this case, the cooling tone is not phase modulated.

A single measurement consists of acquisition of the power spectral density for given system parameters (cooling tone power, detuning, etc.) and determination of the phonon occupancy using Eqs.~\eqref{eq:SII_SI} and~\eqref{eq:nf_SI}, \textit{i.e.} ancillary quantum thermometry and mechanical noise thermometry.
This requires reliable characterizations of $\kappa$, $\bar{n}_c$, $\bar{n}_b$ and $\Delta_c$.
A measurement proceeds as follows.
First, we determine the individual input and reflected powers of the cooling tone and (for two-tone measurements) blue probe by blocking each in turn.
For the two-tone experiments, we nominally set the blue probe power to be a factor of 6 weaker than the cooling tone.
Second, we perform coherent optomechanical spectroscopy to determine $\Delta_c$ and $\kappa$.
Third, we switch to the BHD setup and acquire the power spectral density of the photocurrent with the reflected signal sent to the BHD.
We also take the shot-noise spectra for each measurement by blocking the reflected light from the BHD, to account for the LO power drift across measurements.
Fourth, we record again the total input and reflected probe powers.
The probe powers fluctuate by less than 1\% across measurements.

\section{Data Analysis}
\label{sec:analysis}

\subsection{Coherent optomechanical spectroscopy}

\begin{figure}
	\includegraphics[scale=1]{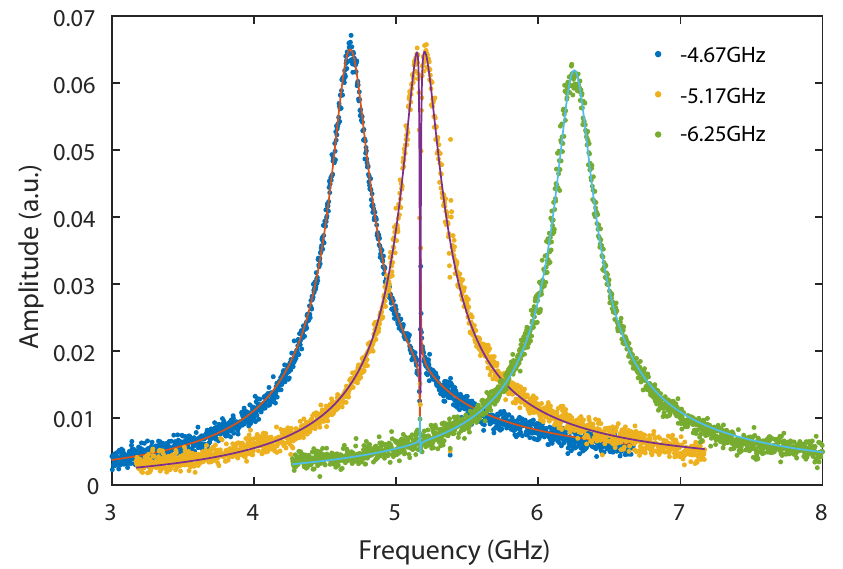}
	\caption{\textbf{Coherent optomechanical spectroscopy.}
Reflected cavity response for single tone detuning sweeps for various detunings of the cooling tone with respect to the optical resonance.
The curve, including the optomechanically-induced transparency, is fitted with a theoretical model to obtain the $\Delta_c$ and $\kappa$.
	}
	\label{fig:OMIT}
\end{figure}

Figure~\ref{fig:OMIT} shows typical coherent optomechanical spectra at several different values of $\Delta_c$ for the single cooling tone measurements with same input powers as in Fig.~3(c) in the main text.
The mechanical motion leads to destructive interference with the probe generated by the phase modulator, resulting in optomechanically induced transparency (OMIT)~\cite{Weis2010,Safavi-Naeini2011} in the reflected cavity response.
We fit the data with a theoretical model described by Ref.~\onlinecite{Safavi-Naeini2011} to extract $\kappa$ and $\Delta_c$, which are used along with the measured powers, to determine the intracavity photon numbers.

\subsection{Calibration using sideband asymmetry}

\begin{figure}
	\includegraphics[scale=1.]{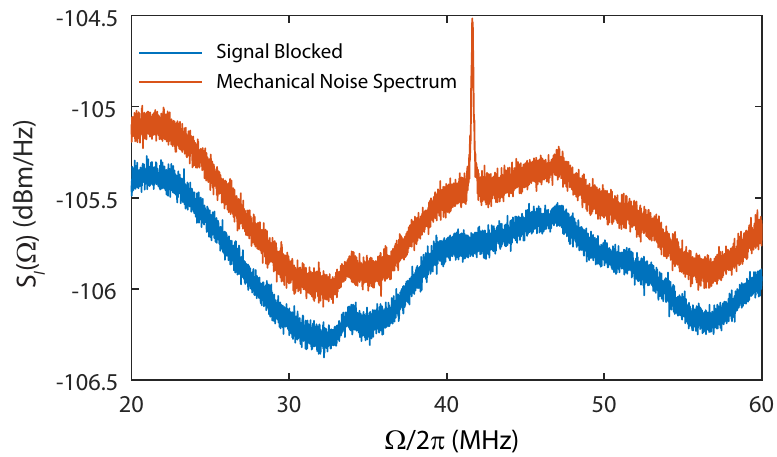}
	\caption{\textbf{Incoherent noise spectrum from heterodyne detection.}
Typical noise spectrum from BHD in the single-tone detuning-sweep measurements at $\Delta_c/2\pi \simeq-7.18\unit{GHz}$. The shot noise spectrum with signal blocked is shown in blue while the noise spectrum of the thermomechanical sideband is shown in red.
}
	\label{fig:spectrum0}
\end{figure}

\begin{figure}
	\includegraphics[scale=1.]{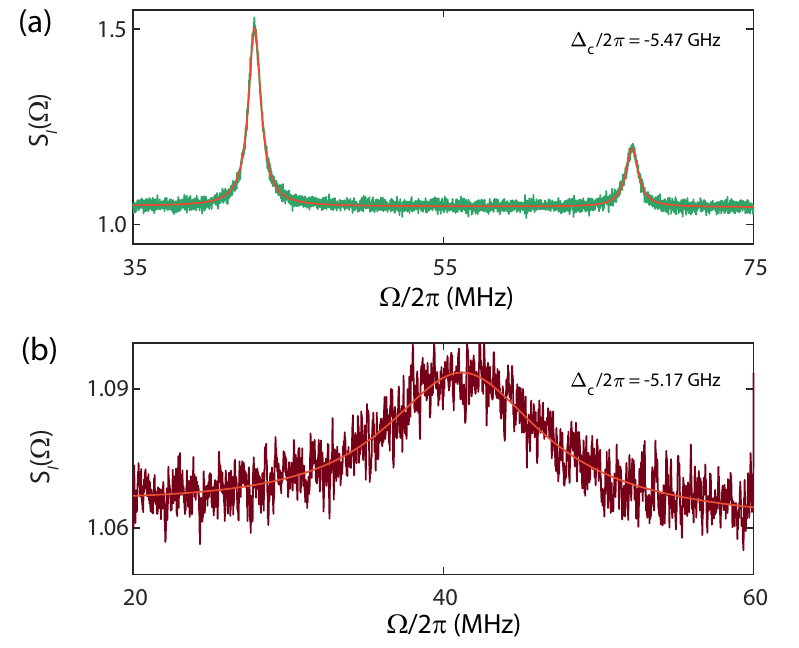}
	\caption{\textbf{Fitting of the incoherent noise spectrum.}
Typical noise spectrum from (a) a two-tone measurement  with $\Delta_c/2\pi\simeq-5.47\unit{GHz}$, and (b) a single-tone measurement with $\Delta_c/2\pi\simeq-5.17\unit{GHz}$, along with the corresponding fitting curve.
}
	\label{fig:spectrum}
\end{figure}

Typical incoherent noise spectra  from the BHD are shown in Fig.~\ref{fig:spectrum0}.
The blue curve corresponds to the shot noise and is obtained by blocking the signal beam in the BHD.
We note that the uneven shot noise floor originates from the frequency dependent gain of the balanced detector.
The red curve corresponds to the thermomechanical noise spectrum in the single-tone detuning-sweep measurements at $\Delta_c/2\pi \simeq -7.18 \unit{GHz}$.
For convenience, we normalize the noise spectrum to the shot noise as shown in Fig.~\ref{fig:spectrum}.
Figure~\ref{fig:spectrum}(a) shows the single-sided noise spectrum consisting of the scattered sidebands from two-tone sideband asymmetry measurements.
Accordingly, we use a fitting function with two Lorentzian terms,
\begin{equation}
\label{eq:Sfit}
\begin{split}
S\s{fit}(\omega) =
c + \frac{\Gamma\s{eff}A_1}{\Gamma\s{eff}^2/4+(\omega-\omega_1)^2}+ \frac{\Gamma\s{eff}A_2}{\Gamma\s{eff}^2/4+(\omega-\omega_2)^2}
\end{split}
\end{equation}
where $c$, $A_1$, $A_2$, $\omega_1$, $\omega_2$ and $\Gamma\s{eff}$ are the fitting parameters.
$c$ corresponds to the noise background.
$A_1$ ($A_2$) and $\omega_1$ ($\omega_2$) correspond to the area and center frequency of the sideband from the cooling tone (blue probe), with effective linewidth $\Gamma\s{eff}$.
From Eq.~\eqref{eq:SII_SI} we have $A_1=\eta\Gamma_c\nf$ and $A_2=\eta\Gamma_b(\nf+1)$,
where $\Gamma_b$ and $\Gamma_c$ are given by Eq.~\eqref{eq:Gamma}.
We can therefore determine both the phonon occupancy
\begin{equation}
\nf = \frac{A_1/\Gamma_c}{A_2/\Gamma_b-A_1/\Gamma_c}.
\end{equation}
and the calibration coefficient
\begin{equation}\label{eq:Cal}
C_{\unit{cal}} = A_2/\Gamma_b-A_1/\Gamma_c,
\end{equation}
which fully calibrates the measurement.

\begin{figure}
  \centering
	\includegraphics[scale=1.0]{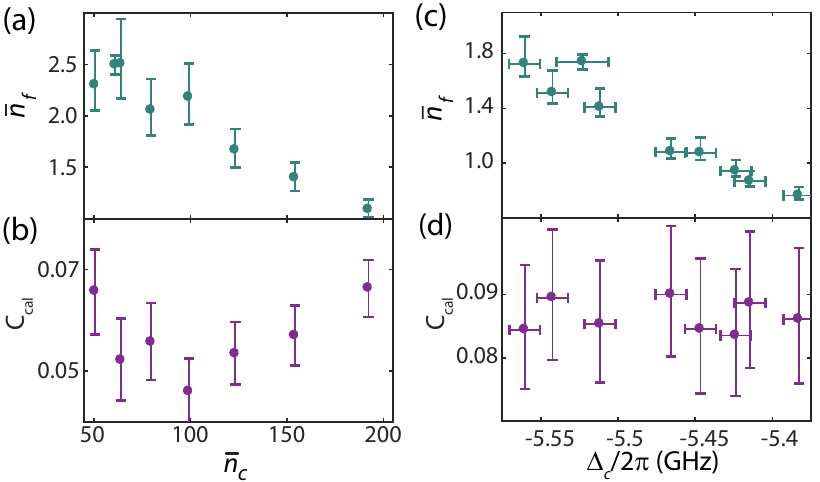}
  \caption{	\textbf{Ancillary Quantum Thermometry for power-sweep and detuning-sweep measurements. }
  (a) and (b), calibrated final occupancy and calibration coefficient ($C_{\unit{cal}}$) from sideband asymmetry of power sweep measurements.   $C_{\unit{cal}}$ is the ratio between the normalized sideband area and the calibrated $\bar{n}_f$.
  The error bars correspond the errors from the Lorentzian fitting of the noise spectrum.
  (c) and (d), calibrated final occupancy and calibration coefficient ($C_{\unit{cal}}$) from sideband asymmetry of detuning sweep measurements. The error bars include the errors in the Lorentzian fitting of the noise spectrum and also the detuning uncertainty of $10\unit{MHz}$.
}
  \label{fig:Cal}
\end{figure}
In Fig.~\ref{fig:Cal}, we show the ancillary quantum thermometry for both the power-sweep and detuning-sweep measurements, including the phonon occupancy and the calibration coefficient.
For the power-sweep measurements as shown in Fig.~\ref{fig:Cal}(a) and~(b), we choose series of pumping powers, which ensure both sufficient laser cooling and measurable while non-overlapping Stokes and anti-Stokes sidebands~\cite{Weinstein2014}.
For the detuning sweep measurements as shown in Fig.~\ref{fig:Cal}(c),(d) we choose series of $\Delta_c$ close to the red sideband to obtain sufficient laser cooling and measurable while non-overlapping Stokes and anti-Stokes sidebands.
The different calibration coefficients between the power-sweep and detuning-sweep measurements are mainly due to  due to the different coupling conditions of the tapered fiber.
The averaged calibration coefficient along with the corresponding standard deviation is used for the ancillary quantum thermometry in the single-tone measurements.

The final occupancy can be determined from the sideband area,
\begin{equation}\label{eq:nfCal}
\nf =\frac{A_s}{\Gamma_s C_{\unit{cal}}},
\end{equation}
where $A_s$ and $\Gamma_s$ are the sideband area and the scattering rate of the cooling tone for the single-tone measurement as shown in Fig.~\ref{fig:spectrum}(b).

\subsection{Mechanical Noise Thermometry}

The occupancy can also be determined using mechanical noise thermometry by anchoring the normalized thermomechanical noise area to the cryogenic thermometer,
\begin{equation}\label{eq:nfkBT}
\nf= \frac{A_s/\Gamma_s}{A_s^0/\Gamma^0_s}\frac{k_B T}{\hbar\Omega_m}\frac{\Gamma_m}{\Gamma^0_s+\Gamma_m}.
\end{equation}
where $A^0_s$ and $\Gamma^0_s$ are the sideband area and the scattering rate of the cooling tone at a specific anchor data point.
In this case, it is assumed that the mechanical mode temperature $T$ is given by the resistive thermometer and there is no excess heating at the anchor point.
The mean phonon occupancy of the mechanical oscillator is $\bar{n}\s{th}\simeq k_B T/\hbar\Omega_m$ when the mechanical mode is in equilibrium with the thermal reservoir.
In the case of negligible optomechanical damping ($\Gamma^0_s\ll\Gamma_m$), Eq.~\eqref{eq:nfkBT} can be simplified to $\nf=\frac{A_s/\Gamma_s}{A_s^0/\Gamma^0_s}\frac{k_B T}{\hbar\Omega_m}$.

\begin{figure}
	\includegraphics[scale=1.0]{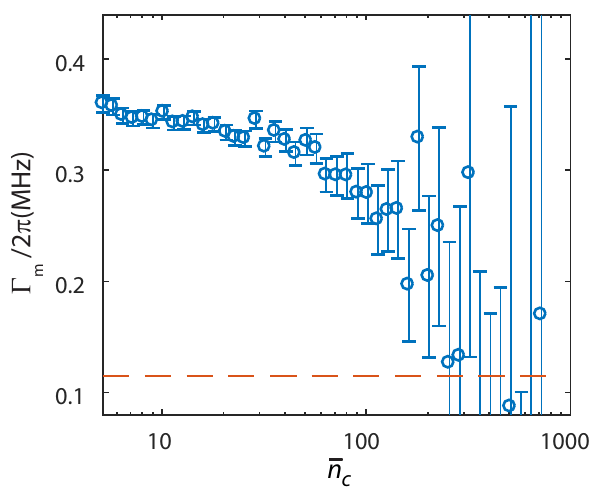}
	\caption{
\textbf{Mechanical Linewidth in the single-tone power-sweep measurements.}
		$\Gamma_m$ is inferred from the fitted $\Gamma\s{eff}$ and calculated $\Gamma_c$ from experimental values. The error bars corresponds to the fitting error in the $\Gamma\s{eff}$. A horizontal dashed line of $\Gamma_m/2\pi = 115\unit{kHz}$ is shown for comparison.
}
	\label{fig:Gammam}
\end{figure}

In the power-sweep series, $\Delta_c=-\Omega_m$, we have $\nf=\frac{A_s/n_c}{A_s^0/n^0_s}\frac{k_B T}{\hbar\Omega_m}$.
We note that knowledge of $\Gamma_m$ is required for the mechanical noise thermometry, as the additional optomechanical damping has to be considered.
From independent single-tone sideband cooling measurements performed at a temperature of $4\unit{K}$ and pressure of $40\unit{mbar}$, we obtain a mechanical damping rate $\Gamma_m/2\pi \sim 115\unit{kHz}$ and vacuum optomechanical coupling rate $g_0/2\pi \sim 1.08\unit{MHz}$.
As evident from the Fig.~2(b) in the main text, the mechanical linewidth is larger  for low input powers. This is due to condensed $^3$He on the sample surface, as the lower temperatures $\sim 2\unit{K}$ in the experiment are obtained by pumping a condensed $^3$He reservoir.
In {Fig.~\ref{fig:Gammam}}, we show the inferred $\Gamma_m=\Gamma\s{\unit{eff}}-\Gamma_c$ for different $\bar{n}_c$, where $\Gamma_c$ is calculated from the experimental values. The error bars correspond to the fitting error in $\Gamma\s{eff}$.
At low pumping powers, the SNR is decreased due to the broadened $\Gamma_m$.
To have a measurable anti-Stokes signal in the noise thermometry shown in Fig.~2, we start from intracavity photon number $\bar{n}_c=5$, with calculated  $\Gamma_c/2\pi = 93\unit{kHz}$ based on experimental values.
From the fitted effective mechanical linewidth $\Gamma\s{eff}$, we estimate $\Gamma_m/2\pi\simeq 360\unit{kHz}$ for $\bar{n}_c=5$.
This is adopted for the mechanical noise thermometry by anchoring at $2\unit{K}$ with $\bar{n}_c=5$ as shown in Fig.~2 in the main text.
For the detuning-sweep series, $\Gamma_m$ is restored to the mechanical linewidth of $2\pi\times 115\unit{kHz}$ at $40\unit{mbar}$ as shown in Fig.~3(d) at large input powers due to the residual optical heating.
The mechanical noise thermometry is anchored at $2\unit{K}$ with $\Delta_c/2\pi\simeq -7.18\unit{GHz}$.
Importantly, the ancilla quantum thermometry is completely independent of $\Gamma_m$ and $\bar{n}\s{th}$.

\begin{figure}
	\includegraphics[scale=1.0]{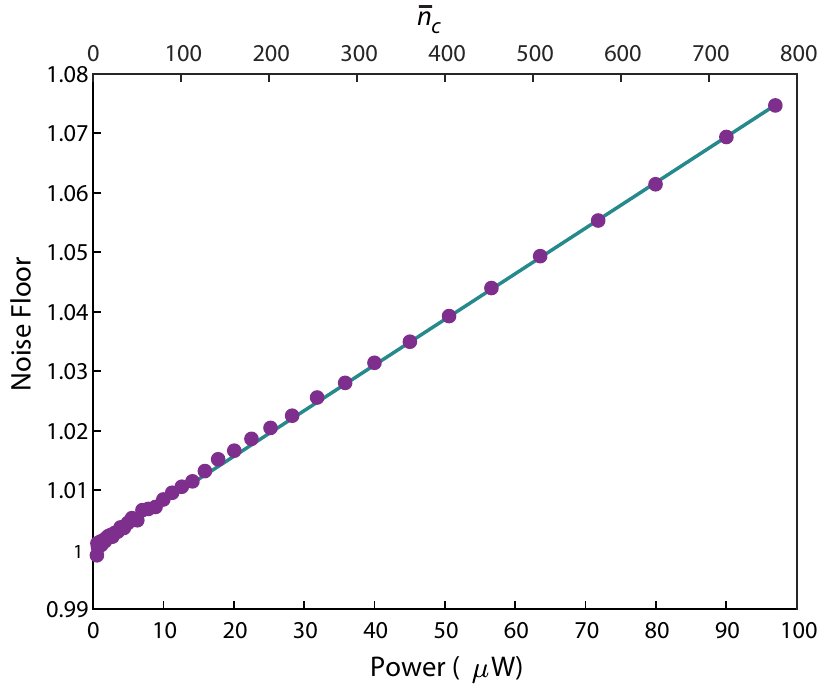}
	\caption{\textbf{Noise floor vs.~reflected power in single-tone power-sweep measurements.}
Noise floor from the normalized noise spectra from the BHD in the single-tone power-sweep measurements (purple full circles), fitted with a green line. The lower x axis corresponds to the reflected power into the balanced heterodyne setup while the upper axis corresponds to the intracavity photon number.
	}
	\label{fig:noisefloor}
\end{figure}

\subsection{Excess laser noise}

Excess laser noise is known to constrain sideband cooling and to corrupt motional sideband asymmetry measurements~\cite{Rabl2009,jayich_cryogenic_2012,safavi-naeini_laser_2013,kippenberg_phase_2013,borkje_heterodyne_2016}.
As shown in our previous work~\cite{qiu_floquet_2019}, the excess laser frequency noise spectrum density $S_{\omega\omega}(\Omega)$ at frequency of $5.2\unit{GHz}$ is measured below $10^5 \unit{rad^2\,Hz}$.
In all the measurements, the pumping tones (cooling tone and blue probe) pass through narrow bandwidth filter cavity locked to the respective pumping tones to reject high frequency excess phase noise.
In Fig.~\ref{fig:noisefloor}, we presented the  noise floor fitted from the thermomechanical noise spectra in the balanced heterodyne measurements (purple full circles), which increases linearly with the reflected power (intracavity photon number) in the single-tone power-sweep measurements (green line).
In BHD, we choose a LO power of around $8\unit{mW}$.
The beating between the highest reflected power ($100\unit{\mu W}$) and vacuum noise from the LO can lead to an increased noise floor by $\sim 1\%$.
The noise floor increase observed in our heterodyne measurements originates from the beating between the high reflected power and excess noise of the LO around $5.17\unit{GHz}$~\cite{qiu_floquet_2019}.
This can be eliminated in principle by passing the LO through a narrow bandwidth filter cavity.
However, this will introduce large insertion loss and experimental complexity, thus is not implemented in our measurements.

\subsection{Error Analysis}
As noted earlier, before and after each set of measurements, both the input powers and the reflected powers of the two tones are checked, and their fluctuation is less than 1\%.
Besides, for each set of measurements the reflection efficiency varies less than 1\%, which eliminates the power/detuning dependence for the calibration efficiency.
We adopt a detuning uncertainty of $\pm 10\unit{MHz}$ for $\Delta_c$ from the fitting error from the coherent optomechanical spectra in both the power-sweep and the detuning-sweep series of measurements.
The detuning uncertainty is taken into consideration for the ancillary quantum thermometry and is included in the error bars of phonon occupancy calibration in addition to the Lorentzian fitting error from the noise spectrum.

\end{document}